\documentclass[10pt,letterpaper]{article}
\usepackage[top=0.85in,left=2.75in,footskip=0.75in]{geometry}

\usepackage{changepage}

\usepackage[utf8]{inputenc}

\usepackage{textcomp,marvosym}

\usepackage{fixltx2e}

\usepackage{amsmath,amssymb}

\usepackage{cite}

\usepackage{nameref,hyperref}

\usepackage{microtype}
\DisableLigatures[f]{encoding = *, family = * }

\usepackage{rotating}

\raggedright
\usepackage[aboveskip=1pt,labelfont=bf,labelsep=period,justification=raggedright,singlelinecheck=off]{caption}

\makeatletter
\renewcommand{\@biblabel}[1]{\quad#1.}
\makeatother

\date{}

\usepackage{lastpage,fancyhdr,graphicx}
\usepackage{epstopdf}
\fancyheadoffset[L]{2.25in}
\fancyfootoffset[L]{2.25in}
\begin{document}
\vspace*{0.35in}

\begin{flushleft}
{\Large
\textbf\newline{The surface age of Sputnik Planum, Pluto,
must be less than 10 million years}
}
\newline
\\
David E. Trilling\textsuperscript{1,2*}
\\
\bigskip
\bf{1} Department of Physics and Astronomy, Northern Arizona
University,
Flagstaff, AZ USA
\\
\bf{2} Lowell Observatory, Flagstaff, AZ USA

%
%





* david.trilling@nau.edu

\end{flushleft}
\section*{Abstract}
Data from the New Horizons mission to Pluto show no
craters on Sputnik Planum down to the detection limit
(2~km for low resolution data, 625~m
for
high resolution data).
The number of small Kuiper Belt Objects that should
be impacting Pluto is known to some degree from 
various astronomical surveys.
We combine these geological and telescopic
observations to make an order of magnitude
estimate that the surface age of Sputnik Planum
must be less than 10~million years.
This maximum surface age is surprisingly
young and implies that
this area of Pluto must be undergoing
active resurfacing, presumably
through some
cryo-geophysical process.
We discuss three possible resurfacing mechanisms
and the implications of each one
for Pluto's physical properties.



\section*{Introduction}
Recent images of Pluto from the New Horizons
spacecraft \cite{science} have revealed a number of surprises.
Chief among these is the complete lack
of detectable craters
in the region
informally known as Sputnik Planum (SP).
(See          
for example
{\tt http://tinyurl.com/ph8bcr5}
and
{\tt http://tinyurl.com/qfto99p} for
low and high resolution images, respectively.)
Several workers \cite{beau,deelia,greenstreet}
made predictions of the expected crater
distribution of Pluto, but none of these
predicted the absence of craters
observed for SP.
We can make an order of magnitude
estimate of the age of this region of
Pluto's surface 
from our knowledge of the population
of small Kuiper Belt Objects (KBOs) in which Pluto orbits.

\section*{Methods}

A number of observatories on the ground
and in space have been used to measure
the size distribution of KBOs,
which is usually expressed as 
$N(<H)$, the number of KBOs
larger than $H$ (typically measured
per square degree).
($H$ is the Solar System
absolute magnitude, which is the magnitude
an object would have 1~AU from the Earth
and 1~AU from the Sun at zero phase;
small values of $H$ correspond to big objects.
Diameter estimates derived from $H$ are
only approximate because the
albedos of the KBOs are not known.)
Two recent results
\cite{parker,fraser2014}
find $N(<H)$ of $10^{1.7}$ for
$H\leq 12$
(around 11~km diameter).
At $H\leq 17$ (around 1~km),
the estimate is 
$10^4$~KBOs
per square degree \cite{B04}.
These results can be extrapolated
to estimate that there are 
$10^{5.8}$~KBOs
larger than $H=22$ (100~meters) per square degree.

The ecliptic plane is something
like 10~degrees in height, so the total
area of the ecliptic is 
360~degrees times 10~degrees, or 3600~deg$^2$.

Pluto sweeps out a torus around the Sun
as it orbits; the volume of this torus is 
$C_{Pl} \pi r^2$, where $C_{Pl}$ is the circumference
of Pluto's orbit (to first order, 40~AU)
and $r$ is Pluto's radius. Gravitational
focusing for Pluto is negligible. The volume
occupied by the Kuiper Belt is
$C_{KB} \pi R^2$, where $C_{KB}$ is
approximately 40~AU, and $R$ is the 
width of the main part of the Kuiper Belt,
around 2~AU. Thus, the fraction of the Kuiper
Belt that Pluto sweeps out is simply
$(r/R)^2$, which is around $1.5\times 10^{-11}$.
This fraction must be reduced by the ratio
of the mean impact velocity (estimated to be around 2~km/sec by \cite{beau})
to
Pluto's orbital velocity of 4.7~km/sec to
account for the fact that Pluto is not orbiting
in a static field of impacting KBOs, but 
rather that the surrounding KBO swarm
is also orbiting the Sun.

The number of impacts onto Pluto per Pluto
year as a function of size (that is, $H$), on average, is therefore given by

\begin{equation}
N(<H)~{\rm deg^{-2}} \times 3600~{\rm deg}^2 \times \frac{2.0~{\rm km/sec}}{4.7~{\rm km/sec}} \times 1.5\times10^{-11}.
\end{equation}

One over this number is therefore the mean
impact interval, in Pluto years; this number
times 250~gives the impact interval in 
Earth years. 

The entirety of SP, which 
covers some 2.5\% of the surface of Pluto, has been 
imaged at the relatively low resolution
of 400~m/pixel. No craters are evident
in this data set, giving a conservative result
that no craters larger than 2~km
exist in SP (using 5~pixels conservatively
as the detection limit).
A small fraction of SP, amounting to
around 0.13\% of the surface of Pluto,
has been imaged at the relatively high
resolution of 125~m/pixel. No craters are
evident in this data set either, giving
a conservative result that no craters
larger than $\sim$625~m
exist in this
small region of SP.
We correct the impact interval above
by this surface coverage fraction
(1/2.5\% and 1/0.13\%, respectively).

Recent detailed work on the cratering behavior
on Pluto \cite{beau} predicts that
$D$, the final crater
diameter, is proportional to $d^{0.783}$, where
$d$ is the impactor diameter.
The 2~km
crater detection limit in the low resolution
data corresponds roughly to 400~m
impactors, while
the 625~m
crater detection limit in the
high resolution imagery corresponds 
roughly to
90~m
impactors.

\section*{Results and discussion}

The imagery requires 
that no craters caused by 400~m (low resolution)
or 90~m (high resolution)
impactors exist in SP. Given the known
impact interval as a function of size (from above),
we can estimate the maximum surface age
of SP.
The results are shown in Fig~\ref{rate}.
The (black, red) lines show the constraints
provided by the (low, high) resolution
images.
In both cases, the conclusion is that
the surface age of SP must be less
than around 10~million years.

This maximum surface age is surprisingly
young and implies that
this area of Pluto must be undergoing
active resurfacing, presumably
through some
cryo-geophysical process.
There are at least three potential
mechanisms by which craters could be erased
from SP.
The following discussion
is largely adapted from \cite{melosh}.

It is possible that craters in SP undergo
{\em viscous relaxation}, in which the surface
material flows to relieve any topographic
features and horizontal and vertical stresses.
The viscous relaxation timescale $\tau_R$ --- the time
it takes for the height of a surface feature to
diminish by a factor of $1/e$ --- is
given approximately by $3 \eta / \rho g w$,
where $\eta$ is the effective viscosity,
$\rho$ is the density, 
$g$ is the gravitational acceleration, and
$w$ is the breadth of the depression (in this case,
a crater).
The density of Pluto's nitrogen ice is
around 1000~kg/m$^3$ \cite{scott}
and 
the gravitational acceleration on Pluto is 
around 0.66~m/s$^2$. To cause a 
625~meter crater (the smallest 
size detectable in the data)
to relax over $10^7$~years
therefore requires an effective viscosity of
the SP surface layer material, which is largely
nitrogen ice,
of around
$4\times10^{19}$~Pa-s.
Because the relaxation timescale is
an upper limit (the surface must be
younger than $10^7$~years), the actual
effective viscosity must be equal to or
less than this value.
This is a relatively loose constraint on viscosity;
a tighter constraint arises
from the next interpretation.

A second possibility is that craters in SP are
erased and the surface reset through {\em convective
overturn}. The physics of this anomaly correction
is similar to that of viscous relaxation, except that
the proximate cause is now a temperature 
difference from the bottom to the top of the 
convective cell. This temperature difference
causes a density anomaly $\Delta \rho$.
The overturn timescale 
$\tau_{overturn}$ is therefore
approximately
$\eta / \Delta \rho g L$, where $L$
is now the vertical dimension.
The difference in density between nitrogen
ice at 40~K (Pluto surface temperature)
and 60~K (the temperature
near the base of the ``cryo-lithosphere''
just below the nitrogen melting temperature)
is around 5\% \cite{scott},
so we use this value for $\Delta \rho$.
We assume
that the 
``cell boundaries'' seen in the images
indicate the horizontal extent of the 
convection cells --- around 30~km ---
and that the vertical size of a convection
cell is around three times smaller than the 
horizontal extent, or around 
10~km.
We find 
a viscosity of equal to or less than
around $10^{17}$ Pa-s.
We have not taken
into account the stress dependence of 
the effective viscosity, which would lower
our estimate somewhat. Nevertheless,
this result is consistent with the viscosity of
nitrogen ice at 45~K derived by \cite{yamashita}
of around $10^8$~Pa-s,
and indicates that convective overturn is 
a plausible mechanism for removing craters
of this size on SP.

A third possible mechanism to erase craters
on SP is through {\em cryovolcanism}
that conveys melt
from a subsurface reservoir. The assumption
here is that at the base of the SP surface layer,
which we again take to be on the order of 10~km,
there is (perhaps partial) melting of nitrogen ice. 
This liquid material, which is under pressure,
could be extruded to the surface
through local cracks (presumably the 
same cell boundaries described above) and fill
in any negative topography before freezing.
This mechanism requires that 
the temperature at the base of the SP surface
layer be around 63~K, at which temperature solid
nitrogen melts, compared to the surface temperature
of 38~K.
This implies a temperature difference
of around 25~K
(the temperature difference across
the SP surface layer) over a vertical
distance of around 10~km, for a thermal
gradient of around 2.5~K/km.
%
%
The volume
of infill material extruded in $10^7$~years
must equal the volume of the crater that is
erased, which is roughly
$\pi D^3 / 80$, where the factor
of~10 in the denominator arises from
the typical depth/diameter ratio of~1:10.
The melt production rate must therefore
be around 1~m$^3$/year or
$10^7$~m$^3$
(0.01~km$^3$)
in ten million years
in order to erase a single crater of
order 625~meters in diameter.

A last uncertainty in the above constraints
is that the size distribution of KBOs
at 100~meters is not well known. In our
estimate here we have used a reasonable
extrapolation from \cite{B04}. However,
the number of KBOs at that size range 
could plausibly be a factor of ten
greater \cite{schlichting},
which makes the surface age of SP younger
by that same factor. Alternately, if 
the number of small KBOs is less than
we assumed (for instance, as in
\cite{fraser}), then the surface
age increases by the same factor.
The estimate used here
is an appropriate middle ground.



\section*{Conclusions and future work}

We have used knowledge of
the Kuiper Belt from telescopic observations
to constrain the age of Sputnik Planum, Pluto.
In the future, additional high-resolution
imaging of SP as well as well-characterized
crater counting on Pluto's surface could be
used to constrain the small size
end of the KBO population. In particular, 
a better understanding 
of the crater detection limit in SP
will help constrain the number
of KBOs smaller than 100~meters.
Alternately, very deep and well-characterized
surveys for small KBOs might place interesting
constraints on the cryo-geophysics of Pluto.

\begin{figure}[h]
\caption{
{\bf Impact interval onto Sputnik Planum, Pluto, in Earth
years as a function of impactor size.}
The black line shows the constraint from
the low resolution imaging (400~m/pixel),
which has greater areal coverage; the red line
shows the constraint from the small amount
of high resolution (125~m/pixel) imaging
available at the time of writing.
The size distribution of KBOs larger than
10~km is taken from \cite{parker}.
For sizes smaller than 10~km, estimates and
extrapolations from \cite{B04}
are used.
Impactors of size 400~m and 90~m
would create craters with the indicated
sizes (as derived from scaling laws presented in
\cite{beau}),
which would be at the resolution limit
of the low and high resolution imaging,
respectively.
We conclude that the surface of Sputnik Planum
must be younger than around 10~million Earth
years. 
This conclusion is drawn both
from the fact that
no craters 2~km in diameter have
been detected in the low resolution
data and no craters 625~meters
in diameter have been detected
in the smaller amount of high resolution imagery.
The two measurements give
the same constraint for the maximum
surface age of Sputnik Planum.
}
\label{rate}
\end{figure}

\section*{Acknowledgments}
I thank Will Grundy
for many helpful conversations and
John Compton for suggestions
about geophysical possibilities.
Two anonymous referees provided
helpful comments.


%
%
%

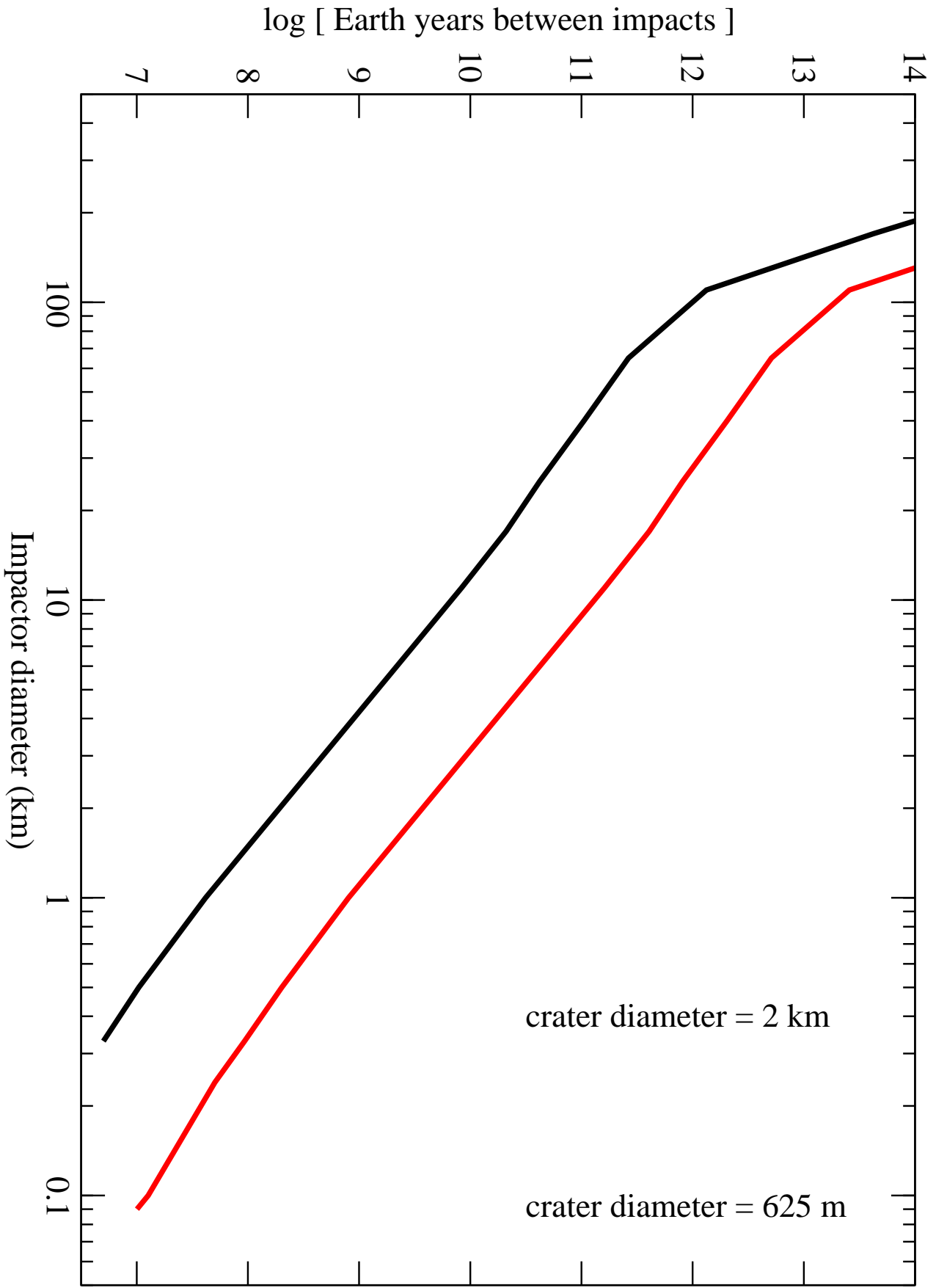
\end{document}